\documentclass[conference]{IEEEtran}
\pdfoutput=1
\usepackage[T2A]{fontenc}
\usepackage[utf8]{inputenc}
\usepackage[russian,english]{babel}

\usepackage{amsthm}
\usepackage{mathrsfs}
\usepackage[ruled,noend]{algorithm2e}
\SetKwInput{kwInit}{Initialization}
\usepackage{tikz,lipsum}
\usepackage[labelformat=simple]{subcaption}
\usepackage{enumitem}
\usepackage{placeins}
\usepackage{float}
\newcommand{\cov}{\normalfont{\text{cov}}}
\newcommand{\vect}[1]{\normalfont{\textbf{#1}}}
\newcommand{\diff}{\dfrac{\partial}{\partial t}}
\DeclareCaptionLabelSeparator{periodspace}{.\quad}
\usepackage[noadjust]{cite}
\usepackage{amsmath,amssymb,amsfonts}
\usepackage{algorithmic}
\usepackage{graphicx}
\usepackage{textcomp}
\usepackage{xcolor}
\def\BibTeX{{\rm B\kern-.05em{\sc i\kern-.025em b}\kern-.08em
    T\kern-.1667em\lower.7ex\hbox{E}\kern-.125emX}}
    
\begin{document}

\title{\textbf{PEDESTRIAN DEAD-RECKONING ALGORITHMS FOR DUAL FOOT-MOUNTED INERTIAL SENSORS}}

\author{
\IEEEauthorblockN{I. A. Chistiakov}
\textit{Huawei Technologies Co. Ltd.}\\
Moscow, Russia \\
chistyakov.ivan@yahoo.com
\and
\IEEEauthorblockN{A. A. Nikulin}
\textit{Huawei Technologies Co. Ltd.}\\
Moscow, Russia \\
nikulin.alexey@huawei.com
\and
\IEEEauthorblockN{I. B. Gartseev}
\textit{Huawei Technologies Co. Ltd.}\\
Moscow, Russia \\
gartseev.ilia@huawei.com
}

\onecolumn

This is a preprint copy that has been accepted for publication in Proceedings of the 26th Saint Petersburg International Conference on Integrated Navigation Systems (ICINS). \\

© 2019 IEEE.  Personal use of this material is permitted. Permission from IEEE must be obtained for all other uses, in any current or future media, including reprinting/republishing this material for advertising or promotional purposes, creating new collective works, for resale or redistribution to servers or lists, or reuse of any copyrighted component of this work in other works.

\twocolumn

\pagebreak

\maketitle

\begin{abstract}
This work proposes algorithms for reconstruction of closed-loop pedestrian trajectories based on two foot-mounted inertial measurement units (IMU). The first proposed algorithm allows calculation of a trajectory using measurements from only one IMU. The second algorithm uses data\footnote{The data used in the article are available for downloading at\\ http://gartseev.ru/projects/mkins2019.} from both foot-mounted IMUs simultaneously. 
Both algorithms are based on the Kalman filter and the assumption that while a foot is on the ground its velocity is supposed to be zero.
Two methods for comparing the obtained trajectories are proposed, advantages and disadvantages of each method are indicated and a way to optimize the computation time is presented. In addition, a method is proposed for constructing one generalized trajectory of human motion based on the trajectories of each leg.
\end{abstract}


\section{Introduction}

Existing methods for recovering pedestrian trajectories using inertial sensor data make it possible to significantly correct the constructed trajectories using filtering algorithms~[1]--[4]. At the same time, accuracy of the results is highly dependent on the attachment point of the measurment units~[5]. In this work, MPU-9250 sensors~[6] are mounted in the area of foot elevation (Fig.~1). This method of attachment allows using the assumption of zero speed of devices at the moments of contact of a foot with a surface, but it does not provide any additional information about orientation of devices in space. Thus, the lack of accuracy of sensors leads to a significant deviation of calculated trajectories from the real ones.

Since inertial methods give acceptable results only on short trajectories (up to three minutes), an additional restriction is introduced to solve the problem: only closed trajectories are considered. Due to this assumption it is possible to use coincidence of the initial and final points and to significantly influence trajectories constructed during the post-processing of data. Since a new measurement is reflected only in the section of the trajectory corresponding to the last step, we propose a smoothing algorithm that corrects the entire calculated curve.

\begin{figure}[htbp]
\centerline{\includegraphics[width=5.72cm]{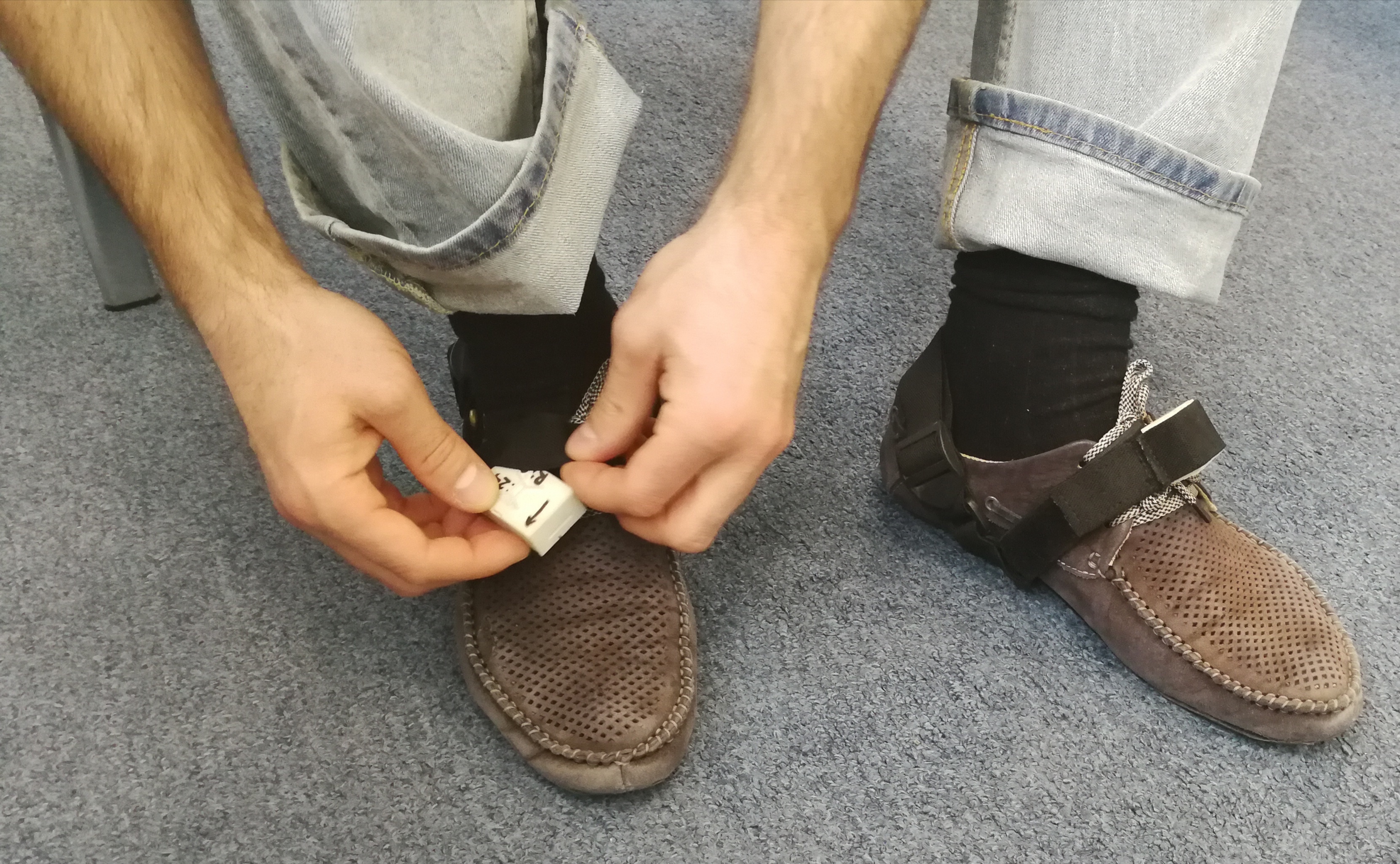}}
\caption{Placement of measurement units.}
\label{IMU_pic}
\end{figure}

Both the position and the orientation of the devices can also be corrected if there are data from two different IMUs fixed respectively on the right and left legs. The advantage of using multiple sensors has been shown, for example, in reference~[7]. In this work, we  use the additional assumption that positions of the sensors in space cannot significantly differ from each other. An algorithm was developed that allows using all of the data mentioned above (information about the initial and final positions, pseudo-observations of velocity, information about position of the other leg) to calculate smooth trajectories of both legs and to construct one generalized trajectory afterwards.

The presented algorithms were tested during experiments with lengths of one to 15 minutes, which took place indoors on horizontal surfaces. In all experiments the devices were mounted in the area of foot elevation.

\section{System description}

\subsection{Non-linear system}

The dynamic system that describes the motion of a foot-mounted IMU is non-linear and has the following form:
\begin{equation}\label{system_C}
    \begin{cases}
    \vect{p}_n = \vect{p}_{n-1} + \vect{v}_{n-1} dt, \\
    \vect{v}_n = \vect{v}_{n-1} + \left(\vect{C}_{n}^T \vect{f}_n + \vect{g}\right) dt, \\
    \vect{C}_n = \vect{R}_n \vect{C}_{n-1},
    \end{cases}
\end{equation}
where $n$ is a time index, $dt$ is a time difference between consecutive measurements, \hbox{$\vect{p}_n \in \mathbb{R}^3$} is the position of an IMU in the navigation frame, \hbox{$\vect{v}_n \in \mathbb{R}^3$} is the velocity vector, \hbox{$\vect{f}_n \in \mathbb{R}^3$} and \hbox{$\vect{w}_n \in \mathbb{R}^3$} are respectively the specific force and angular rate in the body frame, \hbox{$\vect{g}$} is the gravity force, \hbox{$\vect{C}_n \in \mathbb{R}^{3 \times 3}$} is the orientation matrix of the body frame relative to the navigation frame and \hbox{$\vect{R}_n = \vect{R}_n(\vect{w}_n dt)$} is the rotation matrix.

As far as each matrix $\vect{C}_n$ is associated with a set of roll, pitch and yaw angles \hbox{$\boldsymbol{\theta}_n \in \mathbb{R}^3$}, the state vector \hbox{$\vect{x}_n = [\vect{p}_n\ \vect{v}_n\ \boldsymbol{\theta}_n]^T$} is considered.

However, there are other ways of representing system~(1), for example, using quaternions~[1]. In this case, the orientation of the device changes in accordance with the following equation:
\begin{equation*}
\vect{q}_n = \left[ \cos\left( \alpha_n \right) \vect{I}_4 + \dfrac{1}{\alpha_n}  \sin \left( \alpha_n \right) \boldsymbol{\Omega}_n \right] \vect{q}_{n-1},
\end{equation*}
where $\alpha_n = 0.5 \cdot \|\vect{w}_n\| dt$, $\vect{I}_4 \in \mathbb{R}^4$ is the identity matrix,
\begin{equation*}
\boldsymbol{\Omega}_n = 
\dfrac{dt}{2}
\begin{bmatrix}
0 & w_n^3 & -w_n^2 & w_n^1 \\
-w_n^3 & 0 & w_n^1 & w_n^2 \\
w_n^2 & -w_n^1 & 0 & w_n^3 \\
-w_n^1 & -w_n^2 & -w_n^3 & 0
\end{bmatrix},
\end{equation*}
and the velocity vector changes according to
$$
\vect{v}_n = \vect{v}_{n-1} + (\vect{q}_{n-1} \vect{f}_n \vect{q}_{n-1}^* + \vect{g})\ dt.
$$

\subsection{Linearized system}

In the post-processing, we also assume the presence of random measurement errors of accelerometers and gyroscopes ($\tilde{\vect{w}}_n$ and  $\tilde{\vect{f}}_n$ are the measurements):
$$ \vect{w}_n = \tilde{\vect{w}}_n + \delta\vect{w}_n, $$
$$ \vect{f}_n = \tilde{\vect{f}}_n + \delta\vect{f}_n. $$

Let $\vect{a}$ be an arbitrary vector in $\mathbb{R}^3$. We use the following notation:
\begin{equation}\label{hat_notation}
    \hat{\vect{a}} =
    \begin{bmatrix}
        0 & a_3 & -a_2 \\
        -a_3 & 0 & a_1 \\
        a_2 & -a_1 & 0
    \end{bmatrix} \in \mathbb{R}^{3 \times 3}.
\end{equation}

Let $\vect{C} = \vect{C}(t)$ be the real orientation matrix of an IMU and $\tilde{\vect{C}} = \tilde{\vect{C}}(t)$ be the calculated orientation matrix. Since they do not coincide (e.g. due to the initial orientation error), then \hbox{$\vect{C}^T = \vect{M} \tilde{\vect{C}}^T$}, where $\vect{M}$ is some orthogonal matrix. If the difference between $\vect{C}$ and $\tilde{\vect{C}}$ is small, then matrix $\vect{M}$ may be described by a small vector of rotation $\boldsymbol{\beta} \in \mathbb{R}^3$. So,
\begin{equation}\label{var_change}
    \vect{C}^T \approx \left(\vect{I} + \hat{\boldsymbol{\beta}}\right) \tilde{\vect{C}}^T.
\end{equation}
Differentiating both sides of the equality, we get:
\begin{equation}\label{C_diff}
    \diff \vect{C}^T = \left( \diff \hat{\boldsymbol{\beta}} \right) \tilde{\vect{C}}^T + \left(\vect{I} + \hat{\boldsymbol{\beta}}\right) \diff \tilde{\vect{C}}^T.
\end{equation}
Since the matrices $\vect{C}$ and $\tilde{\vect{C}}$ satisfy the Poisson formula
$$
\diff \vect{C} =  \left( \hat{\tilde{\vect{w}}} + \delta\hat{\vect{w}} \right) \vect{С}, \qquad
\diff \tilde{\vect{C}} = \hat{\tilde{\vect{w}}} \tilde{\vect{С}},
$$
then~(4) can be rewritten using~(3):
\begin{equation}
\diff \hat{\boldsymbol{\beta}} = -\vect{C}^T \cdot \delta\hat{\vect{w}} \cdot \tilde{\vect{C}},
\end{equation}
\begin{equation}\label{diff_beta}
    \diff \boldsymbol{\beta} \approx -\tilde{\vect{C}}^T \cdot \delta \vect{w}.
\end{equation}

Since the equality $\hat{\vect{a}} \vect{b} = -\hat{\vect{b}} \vect{a}$ is correct for any vectors $\vect{a} \in \mathbb{R}^3$, $\vect{b} \in \mathbb{R}^3$, we get the following equality for $\vect{v}$:
\begin{multline}\label{diff_v}
    \diff \vect{v} = \vect{С}^T \vect{f} + \vect{g} = (\vect{I} + \hat{\boldsymbol{\beta}}) \tilde{\vect{С}}^T \vect{f} + \vect{g} = \hat{\boldsymbol{\beta}} \left( \tilde{\vect{С}}^T \vect{f} \right) + \\
    + \left( \tilde{\vect{С}}^T \vect{f} + \vect{g} \right) =
    -\left( \widehat{ \tilde{\vect{С}}^T \vect{f}} \right) \boldsymbol{\beta} + \left( \tilde{\vect{С}}^T \vect{f} + \vect{g} \right) \approx \\
    \approx -\left( \widehat{ \tilde{\vect{С}}^T \tilde{\vect{f}}} \right) \boldsymbol{\beta} + \left( \tilde{\vect{С}}^T \tilde{\vect{f}} + \vect{g} \right) + \tilde{\vect{C}}^T\delta\vect{f}.
\end{multline}

Using~(6) and~(7), it is possible to change the current set of variables to $\vect{x}_n = [\vect{p}_n\ \vect{v}_n\ \boldsymbol{\beta}_n]^T$ and to consider the linear system with matrix $\vect{F}_n$:
\begin{equation}\label{system_linear}
	\vect{F}_n = 
    \begin{bmatrix}
        \vect{I}_{3 \times 3} & \vect{I}_{3 \times 3} dt & \vect{O}_{3 \times 3} \\
        \vect{O}_{3 \times 3} & \vect{I}_{3 \times 3} & -\widehat{\tilde{\vect{С}}_n^T \tilde{\vect{f}}_n} dt \\
        \vect{O}_{3 \times 3} & \vect{O}_{3 \times 3} & \vect{I}_{3 \times 3}
    \end{bmatrix}.
\end{equation}

\section{Algorithm for one IMU}

The algorithm is based on the use of function $T$ that detects a stationary foot position when walking~[2]. We assume that position is stationary if the value of $T$ is less than some given constant:
\begin{equation}\label{zupt_statistic}
T\left(\left\{\tilde{\vect{f}}_i, \tilde{\vect{w}}_i \right\}_{W_n}\right) < \gamma,
\end{equation}
where $W_n$ is a time window of the length $N$ samples centered around $t_n$, $\gamma > 0$ is a zero-velocity detection threshold.

When the system is stationary, the velocity coordinates of a calculated vector $\vect{x}_n$ can be treated as a pseudo-measurement of the velocity estimation error (since the real velocity is supposed to be zero). Hence, we get observation $\vect{v}_n = \vect{0}$ with observation matrix
$$
\vect{H} =
\begin{bmatrix}
\vect{O}_{3 \times 3} & \vect{I}_{3 \times 3} & \vect{O}_{3 \times 3} \\
\end{bmatrix} \in \mathbb{R}^{3 \times 9}.
$$

The corresponding algorithm~[1] is shown in Algorithm~1. The following notation is used: $\vect{F}_n \in \mathbb{R}^{9 \times 9}$ is the matrix of linearized system~(8), $\vect{P}_n \in \mathbb{R}^{9 \times 9}$ is the error covariance matrix, $\vect{Q} \in \mathbb{R}^{6 \times 6}$ is the covariance matrix of the measurement noise, $\vect{R} \in \mathbb{R}^{3 \times 3}$ is the covariance matrix for the noises of the observations, $\vect{G}_n$ is the process noise matrix:
$$
\vect{G}_n =
\begin{bmatrix}
\vect{O}_{3 \times 3} & \vect{O}_{3 \times 3} \\
\tilde{\vect{C}}_n^T dt & \vect{O}_{3 \times 3} \\
\vect{O}_{3 \times 3} & -\tilde{\vect{C}}_n^T dt
\end{bmatrix}
\in \mathbb{R}^{9 \times 6},
$$
$\tilde{\vect{C}}_n \in \mathbb{R}^{3 \times 3}$ is a calculated estimation of orientation matrix and the function $f_{mech}$ corresponds to formulas~(1).

\begin{algorithm}[ht]
\SetAlgoLined
\DontPrintSemicolon
\SetNoFillComment
\kwInit{$\tilde{\vect{x}}_0 \gets E[\vect{x}_0]$,\ $\vect{P}_0 \gets \cov(\vect{x}_0)$}
\For{$n = 1$ \KwTo $N$}{
    \tcc*[l]{Time update}
    $\tilde{\vect{x}}_n \gets f_{mech}(\tilde{\vect{x}}_{n-1}, \tilde{\vect{f}}_n, \tilde{\vect{w}}_n)$ \\
    $\vect{P}_n \gets \vect{F}_n \vect{P}_{n-1} \vect{F}_n^T + \vect{G}_n \vect{Q} \vect{G}_n^T$ \\
    \tcc*[l]{Measurement update}
    \If{$T\left(\{\tilde{\vect{f}}_i, \tilde{\vect{w}}_i\}_{W_n}\right) < \gamma$}{
        $\vect{K}_n \gets \vect{P}_n \vect{H}^T \left( \vect{H} \vect{P}_n \vect{H}^T + \vect{R} \right)^{-1}$ \\
        $\delta \tilde{\vect{x}}_n \gets \vect{K}_n \tilde{\vect{v}}_n$ \\
        $\vect{P}_n \gets \left( \vect{I}_{9 \times 9} - \vect{K}_n \vect{H} \right) \vect{P}_n$ \\
        \tcc*[l]{Compensate internal states}
        \begin{math}
        \begin{bmatrix}
        \tilde{\vect{p}}_n \\ \tilde{\vect{v}}_n
        \end{bmatrix}
        \gets
        \begin{bmatrix}
        \tilde{\vect{p}}_n \\ \tilde{\vect{v}}_n
        \end{bmatrix}
        +
        \begin{bmatrix}
        \delta \tilde{\vect{p}}_n \\ \delta \tilde{\vect{v}}_n
        \end{bmatrix}
        \end{math} \\
        $\tilde{\vect{C}}_n \gets (\vect{I}_{3 \times 3} + \hat{\boldsymbol{\beta}}_n) \tilde{\vect{C}}_n$ \\
        $\delta \tilde{\vect{x}}_n \gets \vect{0}$,\ \ $\boldsymbol{\beta}_n \gets \vect{0}$
    }
    \caption{Kalman filter.}
}
\end{algorithm}

We also assume that at the beginning and at the end of each experiment measurement units are retained in a resting state for a few seconds. Thus, in these moments it is possible to use additional information about the initial position $\vect{p}_n = \vect{0}$ (without loss of generality we consider it to be zero). Then the observation matrix is
\begin{equation*}
    \vect{H} = 
    \begin{bmatrix}
        \vect{I}_{3 \times 3} & \vect{O}_{3 \times 3} & \vect{O}_{3 \times 3} \\
        \vect{O}_{3 \times 3} & \vect{I}_{3 \times 3} & \vect{O}_{3 \times 3}
    \end{bmatrix} \in \mathbb{R}^{6 \times 9}.
\end{equation*}

In order for the added position adjustment to take place not only at the end of the movement, but to reflect on the whole trajectory, a smoothing RTS filter (Algorithm~2), also described in~[1], is used. Smoothing also allows to get rid of the discontinuities that occur during the correction at the end of each step. However, since the system of equations~(1) is non-linear, the errors of angle estimation increase quickly. That leads to a discrepancy between equations~(1) and linear approximation~(8). To solve this problem, it is suggested to correct the angle values at the forward Kalman filter stage.
The pseudo-code of the proposed algorithm is given in Algorithm~3.

\begin{algorithm}\label{RTS}
\SetAlgoLined
\For{$n = N - 1$ \KwTo $1$}{
		$\vect{A}_n \gets \vect{P}_{n|n} \vect{F}_n^T \vect{P}_{n+1|n}^{-1}$ \\
	    $\tilde{\vect{x}}_{n|N} \gets \tilde{\vect{x}}_{n|n} + \vect{A}_n \left( \tilde{\vect{x}}_{n+1|N} - \tilde{\vect{x}}_{n+1|n} \right) $ \\
	    $\vect{P}_{n|N} \gets \vect{P}_{n|n} + \vect{A}_n \left( \vect{P}_{n+1|N} - \vect{P}_{n+1|n} \right) \vect{A}_n^T$
}
\caption{RTS smoothing.}
\end{algorithm}

The difference between two methods is shown in Fig.~2. Red color path corresponds to the trajectory that was calculated without preliminary angle correction. Yellow color path corresponds to the trajectory that was calculated with the proposed method.
It is known that the first and the last segments of the real trajectory passed along the same line, however, a few meters difference may be noticed in the first case.

\begin{algorithm}\label{advanced_filter}
\SetAlgoLined
\DontPrintSemicolon
\SetNoFillComment
\kwInit{$\tilde{\vect{x}}_0 \gets E[\vect{x}_0]$,\ $\delta \tilde{\vect{x}}_0 \gets \vect{0}$,\ $\vect{P}_0 \gets \cov(\vect{x}_0)$}
\tcc*[l]{Forward Kalman filter stage}
\For{$n = 2$ \KwTo $N$}{
    \tcc*[l]{Time update}
    $\tilde{\vect{x}}_n \gets f_{mech}(\tilde{\vect{x}}_{n-1}, \tilde{\vect{f}}_n, \tilde{\vect{w}}_n)$ \\
    $\delta \tilde{\vect{x}}_{n|n-1} \gets \vect{F}_n \delta \tilde{\vect{x}}_{n-1|n-1}$ \\
    $\vect{P}_{n|n-1} \gets \vect{F}_n \vect{P}_{n-1|n-1} \vect{F}_n^T + \vect{G}_n \vect{Q} \vect{G}_n^T$ \\
    \tcc*[l]{Measurement update}
    \If{$T\left(\{\tilde{\vect{f}}_i, \tilde{\vect{w}}_i\}_{W_n}\right) < \gamma$}{
        \uIf{\normalfont{standstill($n$) = true}}{
            \begin{math}
            \vect{H} \gets
            \begin{bmatrix}
            \vect{I}_{3 \times 3} & \vect{O}_{3 \times 3} & \vect{O}_{3 \times 3} \\
            \vect{O}_{3 \times 3} & \vect{I}_{3 \times 3} & \vect{O}_{3 \times 3}
            \end{bmatrix}
            \end{math}\\
            $\vect{K}_n \gets \vect{P}_{n|n-1} \vect{H}^T \left( \vect{H}\vect{P}_{n|n-1} \vect{H}^T + \vect{R}' \right)^{-1}$ \\
        $\delta \vect{x}_{n|n} \gets \delta \vect{x}_{n|n-1} - \vect{K}_n
        \begin{bmatrix}
        \delta \vect{p}_{n|n-1} - \vect{p}_n \\
        \delta \vect{v}_{n|n-1} - \vect{v}_n
        \end{bmatrix}$
        }
        \Else{
            \begin{math}
            \vect{H} \gets
            \begin{bmatrix}
            \vect{O}_{3 \times 3} & \vect{I}_{3 \times 3} & \vect{O}_{3 \times 3}
            \end{bmatrix}
            \end{math}\\
            $\vect{K}_n \gets \vect{P}_{n|n-1} \vect{H}^T \left( \vect{H}\vect{P}_{n|n-1} \vect{H}^T + \vect{R}'' \right)^{-1}$ \\
        $\delta \vect{x}_{n|n} \gets \delta \vect{x}_{n|n-1} - \vect{K}_n (\delta \vect{v}_{n|n-1} - \vect{v}_n)$ \\
        }
        $\vect{P}_{n|n} \gets \left(\vect{I}_{9 \times 9} - \vect{K}_n \vect{H} \right) \vect{P}_{n|n-1}$ \\
        \tcc*[l]{Compensate internal angle states}
        $\tilde{\vect{C}}_n \gets (\vect{I}_{3 \times 3} + \hat{\boldsymbol{\beta}}_n) \tilde{\vect{C}}_n$ \\
        $\delta \tilde{\boldsymbol{\theta}}_n \gets \vect{0}$,\ \ $\boldsymbol{\beta}_n \gets \vect{0}$
    }
}
\tcc*[l]{Smoothing}
\For{$n = N - 1$ \KwTo 1}{
    $\vect{A}_n \gets \vect{P}_{n|n} \vect{F}^T \vect{P}_{n+1|n}^{-1}$ \\
    $\delta \tilde{\vect{x}}_{n|N} \gets \delta \tilde{\vect{x}}_{n|n} + \vect{A}_n (\delta \tilde{\vect{x}}_{n+1|N} - \delta \tilde{\vect{x}}_{n+1|n})$ \\
    $\vect{P}_{n|N} \gets \vect{P}_{n|n} + \vect{A}_n (\vect{P}_{n+1|N} - \vect{P}_{n+1|n}) \vect{A}_n^T$ \\
}
\tcc*[l]{Compensate internal states}
\For{$n = 1$ \KwTo $N$}{
    $\tilde{\vect{x}}_n \gets \tilde{\vect{x}}_n + \delta\tilde{\vect{x}}_{n|N}$ \\
    $\delta \tilde{\vect{x}}_n \gets \vect{0}$ \\
}
\caption{Algorithm that takes closureness into account.}
\end{algorithm}

\begin{figure}[htbp]
\centerline{\includegraphics[width=9cm]{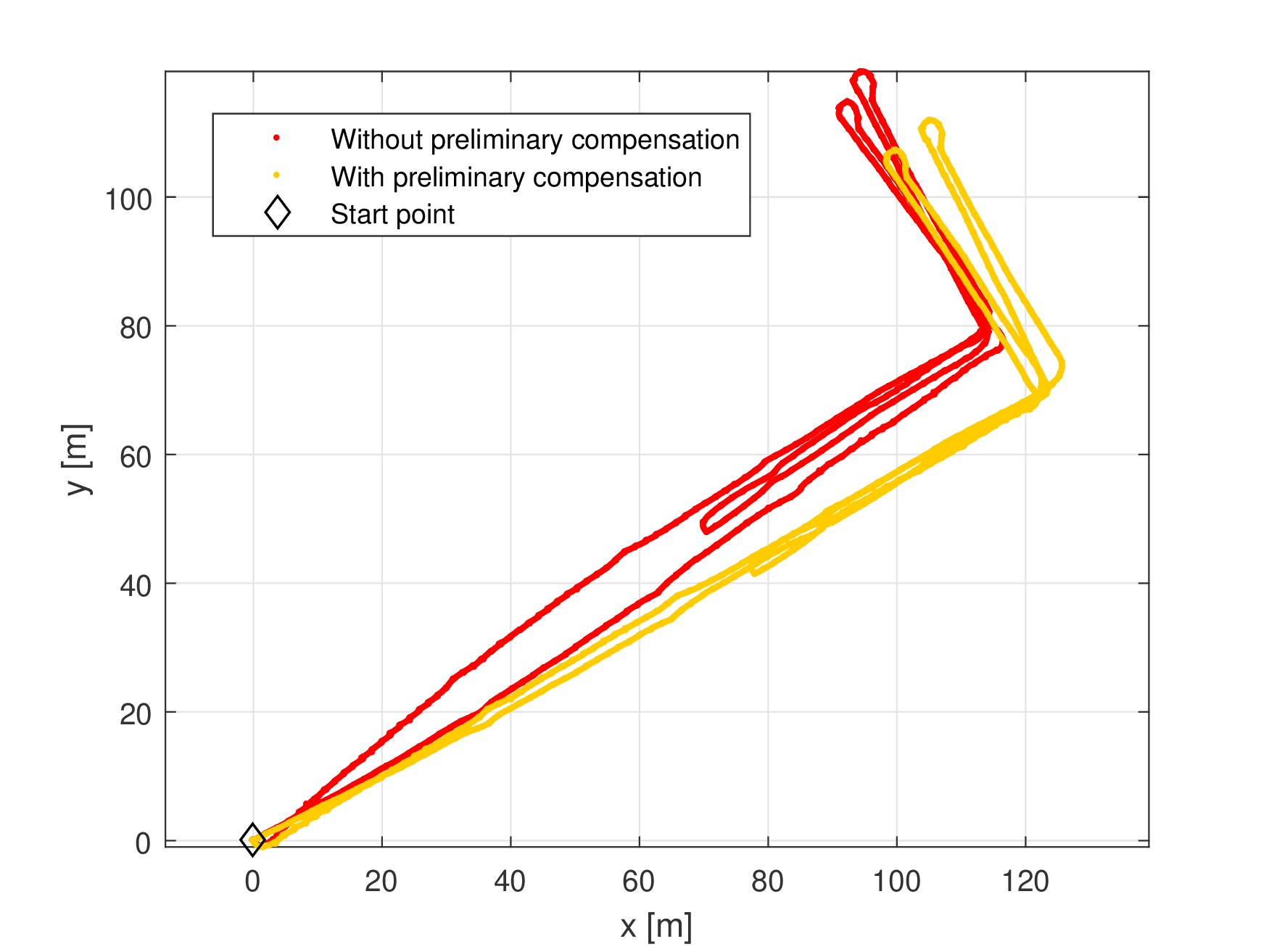}}
\caption{Comparison of trajectories with preliminary correction of angles and without it.}
\label{comp_diff}
\end{figure}

\subsection{Metrics of the path similarity}

To assess the quality of results of the proposed algorithm, it is possible to conduct an experiment in which IMUs are fixed on both legs of a person and then compare two reconstructed trajectories between themselves. To do this, a metric is needed that is capable of comparing two curves. The same metric is needed in order to put optimally one trajectory over another (since the initial orientations of IMUs do not coincide). In this case, the metric should take into account the following features: generally speaking, the durations of the experiment for the left and right legs are different (the sensors are not always switched on at the same time), IMUs do not work synchronously and some measurements may be omitted. Two metrics that satisfy the given requirements were investigated: discrete Fr\'echet distance and DTW-metric.

\subsubsection{Discrete Fr\'echet distance}

Due to the discrete nature of measurements, each calculated trajectory may be treated as a polygonal curve \hbox{$C: [1, n] \to \mathbb{R}^3$}, where $n$ is the number of measurements taken during an experiment. The polygonal curve is uniquely represented with a set \hbox{$\sigma(C) = (C(0), C(1), \ldots, C(n))$} of segment ends, so that for any $\lambda \in [0, 1]$ the following equality is fulfilled:
$$ C(i + \lambda) = (1 - \lambda) C(i) + \lambda C(i + 1),\ \ i = \overline{1, n}. $$

Let $C = (\nu_1,\ldots,\nu_n)$ and $R = (\mu_1,\ldots,\mu_m)$ be two polygonal curves and let $L$ denote the sequence of distinct pairs $(\nu_{a_1}, \mu_{b_1}), \ldots, (\nu_{a_s}, \mu_{b_s})$, where $a_1 = 1$, $b_1 = 1$, $a_s = n$, $b_s = m$ and either $a_i = a_{i-1}$ or $a_i = a_{i - 1} + 1$ is correct for any $i = \overline{2,n}$, as well as $b_j = b_{j-1}$ or $b_j = b_{j - 1} + 1$ is correct for any $j = \overline{2,m}$. The length $\| L \|$ of coupling $L$ is defined as the length of the longest link in $L$:
$$ \|L\| = \max\limits_{1 \le i \le s} d(\nu_{a_i}, \mu_{b_i}), $$
where $d$ is some metric of the $\mathbb{R}^3$ space (e.g. the Euclidean metric).
Discrete Fr\'echet distance is defined to be
$$ F(C, R) = \min\limits_L \{ \| L \| \}. $$

The algorithm developed by Eiter and Manilla in~[8] allows to calculate the discrete Frechet distance over time $O(nm)$ using dynamic programming method, where $m$ and $n$ are the lengths of polygonal curves.

\subsubsection{Dynamic time warping}

The dynamic time warping algorithm (DTW) is a method for calculating an optimal match between two given sequences with certain restrictions:
\begin{itemize}
    \item Every index from the first sequence must be matched with one or more indices from the other sequence, and vice versa.
    \item The first index from the first sequence must be matched with the first index from the other sequence (but it does not have to be its only match).
    \item The last index from the first sequence must be matched with the last index from the other sequence (but it does not have to be its only match).
    \item The mapping of the indices of the first sequence to the indices of the other sequence must be monotonically increasing, i.e. if $j > i$ are indices from the first sequence, then there must not be two indices $l > k$ in the other sequence such that index $i$ is matched with index $l$ and index $j$ is matched with index $k$, and vice versa.
\end{itemize}
The optimal match is denoted by the match that satisfies all the restrictions and that has the minimal sum of distances between the corresponding points.

Let $Q = (q_1, q_2, \ldots, q_n)$ and $C = (c_1, c_2, \ldots, c_m)$ be two time series.
The trivial DTW algorithm consists of the following steps:
\begin{enumerate}[label=\arabic*.]
    \item Calculation of such a matrix $d \in \mathbb{R}^{n \times m}$ that each element $d_{i,j}$ of $d$ is equal to the distance between $q_i$ and~$c_j$.
    \item Calculation of a transformation matrix $D$. Each matrix element is defined as
    \begin{equation*}
        D_{i,j} = 
        \begin{cases}
            d_{1,1}, \text{\ if\ } i = 1, j = 1, \\
            d_{i,j} + D_{i-1, j}, \text{\ if\ }   i \neq 1, j = 1, \\
            d_{i,j} +D_{i, j-1}, \text{\ if\ }  i = 1, j \neq 1, \\
            d_{i,j} + \min \left\{ D_{i-1, j}, D_{i-1, j-1}, D_{i, j-1} \right\}, \\
            \qquad \qquad \qquad \qquad \text{\ if\ }   i \neq 1, j \neq 1.
        \end{cases}
    \end{equation*}
    \item Calculation of the optimal transformation path denoted
    $$ W = (w_1, w_2, \ldots, w_k), $$
    which is a set of adjacent elements of matrix $D$ and which establishes a correspondence between $Q$ and~$C$. Here, $k$ is a number of elements in the sequence~$W$.
    The transformation path is chosen in such a way that the sum of distances between the corresponding points is minimal.
    \item The DTW distance is calculated as follows:
    \begin{equation*}
        DTW(Q, C) = \dfrac{1}{k} \sum\limits_{i=1}^k d(w_i).
    \end{equation*}
\end{enumerate}

The time complexity of the presented algorithm is $O(nm)$. However, it may be reduced by using one of the algorithm modifications~[9].

\subsection{Comparison of the metrics}

\begin{figure}[htbp]
\begin{subfigure}[b]{\linewidth}
\centering%
\centerline{\includegraphics[width=9cm]{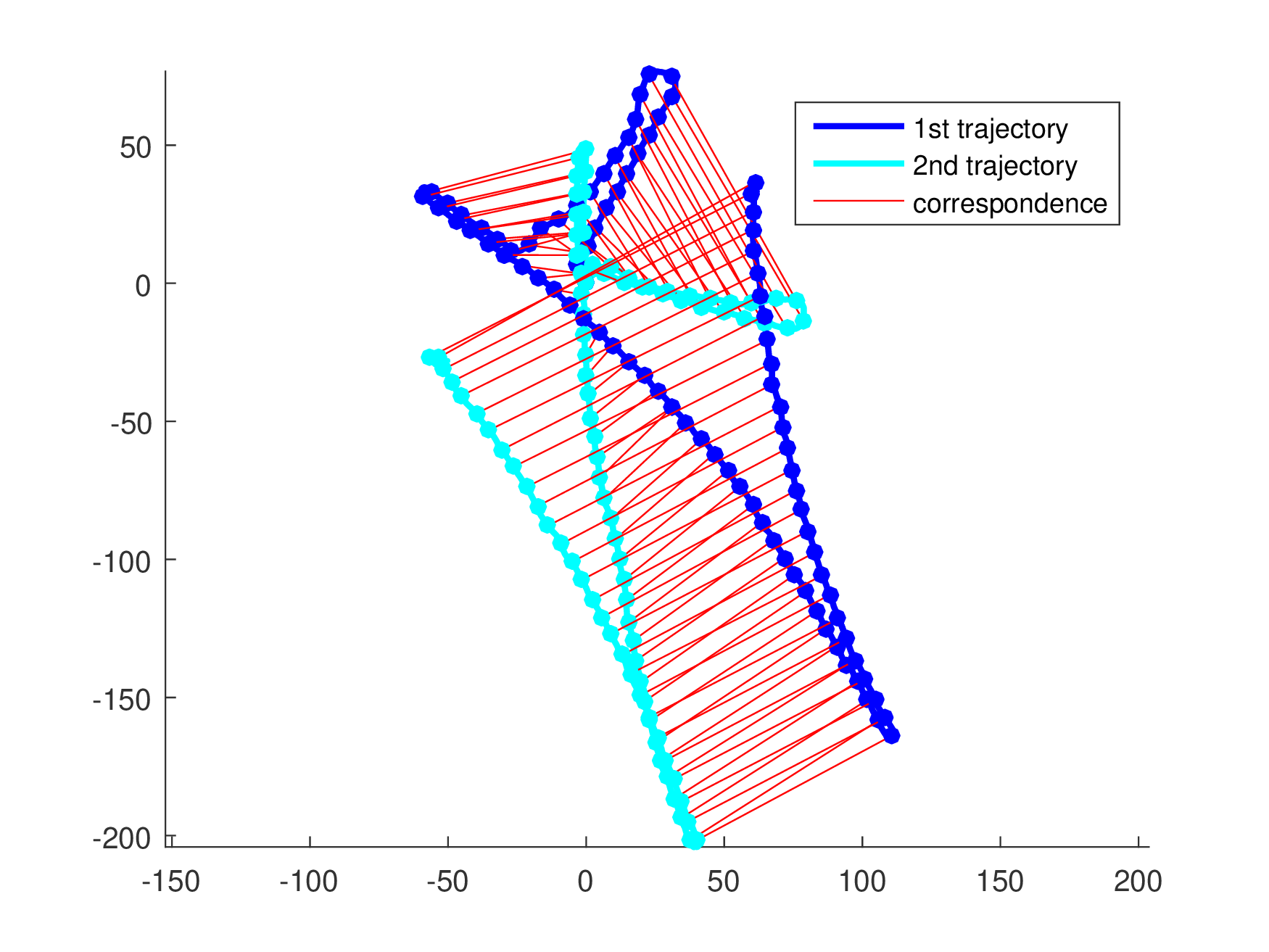}}
\caption{DTW-metric}
\end{subfigure}\\
\begin{subfigure}[b]{\linewidth}
\centering%
\centerline{\includegraphics[width=9cm]{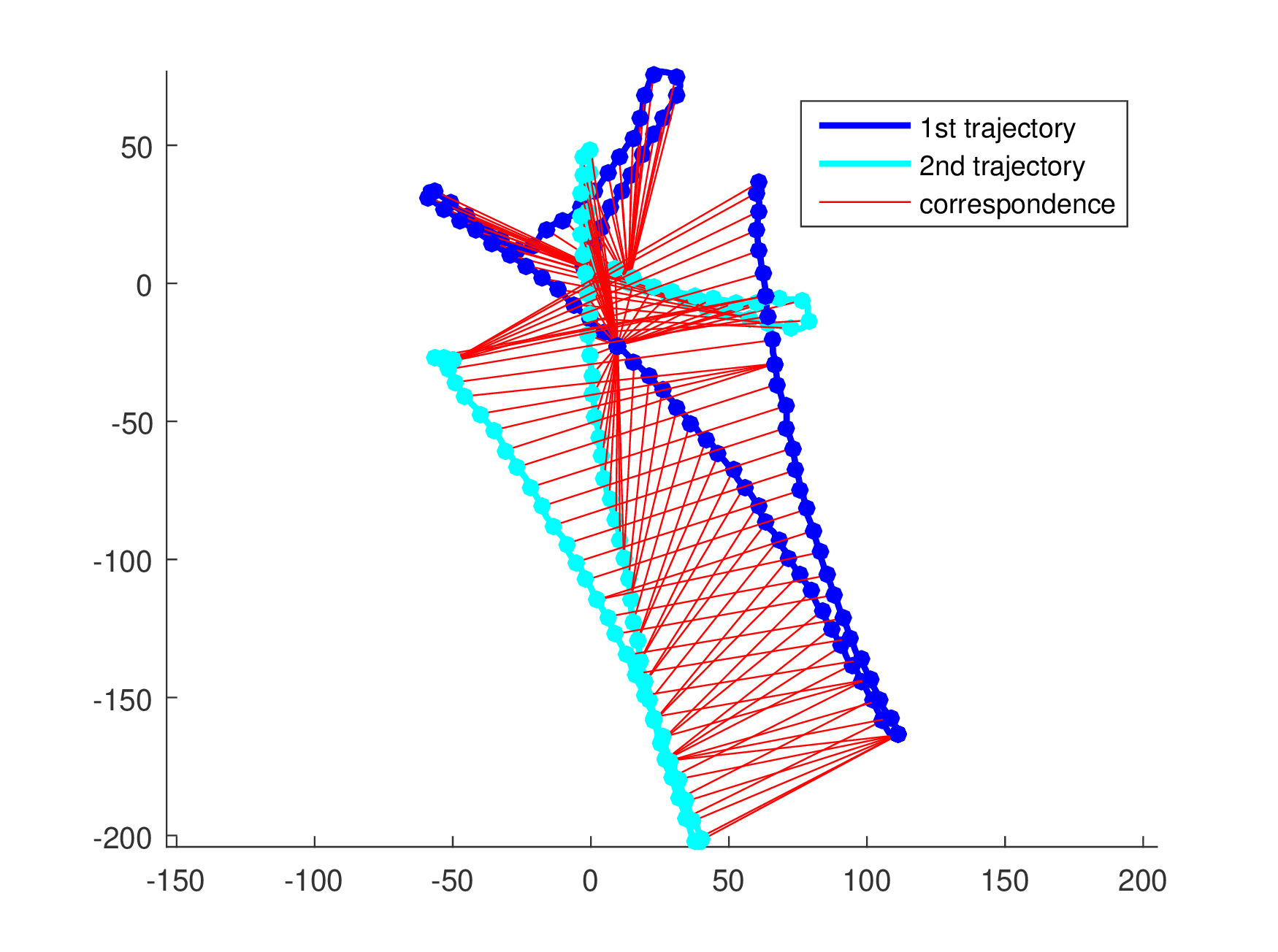}}
\caption{Fr\'echet metric}
\end{subfigure}
\captionsetup{justification=centering}
\caption{Comparison of the metrics}\label{8-40}
\end{figure}

For comparison of the metrics, trajectories of different legs were analyzed and pairs of corresponding points were selected.

It is shown in Fig.~3 that the DTW algorithm selects pairs of points more accurately (some matches are marked in red). Fr\'echet metric gives worse results in cases when the corresponding parts of the curves are located at a large angle to each other. Therefore, further proposed algorithms use DTW-metric.

Nevertheless, there is an example (see Fig.~4) when both algorithms fail to detect correspondence correctly. Incorrect mapping takes place when turning points of trajectories are at a significant distance from each other.

\begin{figure}
\begin{subfigure}[b]{\linewidth}
\centering%
\centerline{\includegraphics[width=9cm]{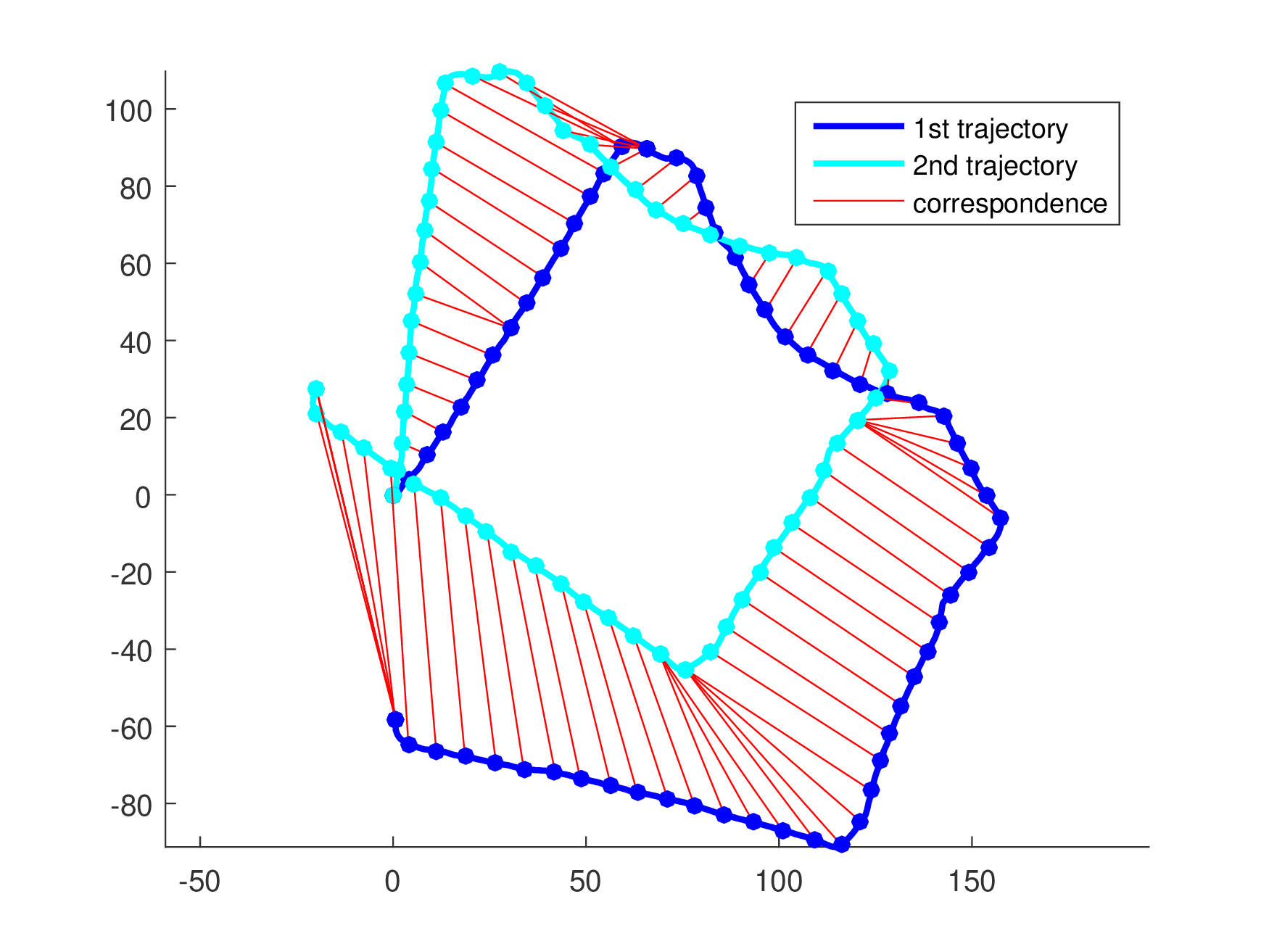}}
\caption{DTW-metric}
\end{subfigure}\\
\begin{subfigure}[b]{\linewidth}
\centering%
\centerline{\includegraphics[width=9cm]{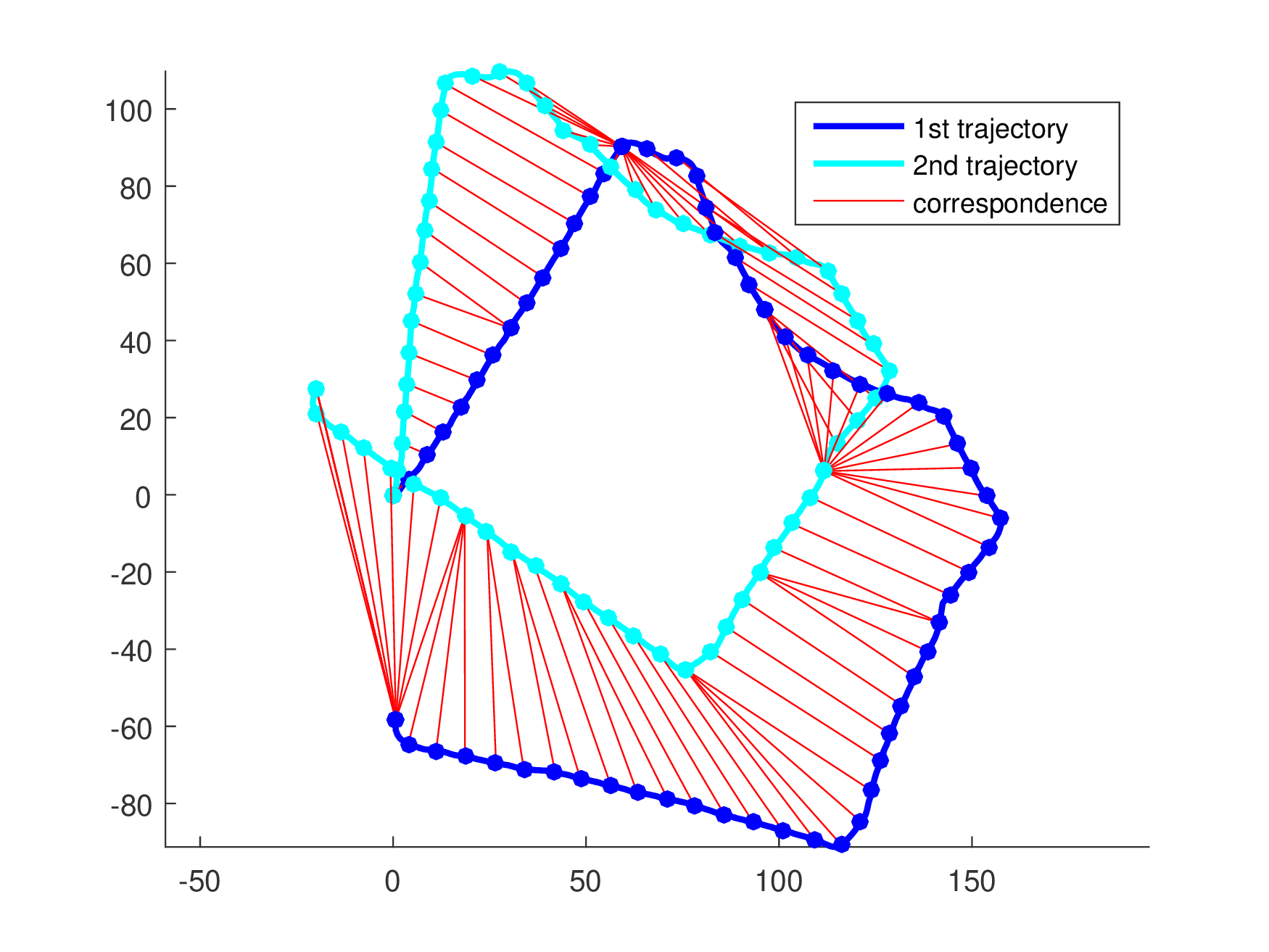}}
\caption{Fr\'echet metric}
\end{subfigure}
\captionsetup{justification=centering}
\caption{Comparison of metrics without the use of additional constraint.}\label{8-158-inf}
\end{figure}

In some cases points that are far apart in time can be mapped by an algorithm although such mapping is undesirable. Thus, the results of the algorithms can be improved by using an additional parameter $w$~--- the maximum allowed difference between the indices of the corresponding points (index is the ordinal number of a measurement in the current trajectory). It is desirable to choose a value $w$ which does not exceed the average number of measurements during one step. The result of using this approach is shown in Fig.~5. If two sequences have the same number of points and $w = 1$, then both methods match points with equal sequence number.

\begin{figure}
\begin{subfigure}[b]{\linewidth}
\centering%
\centerline{\includegraphics[width=9cm]{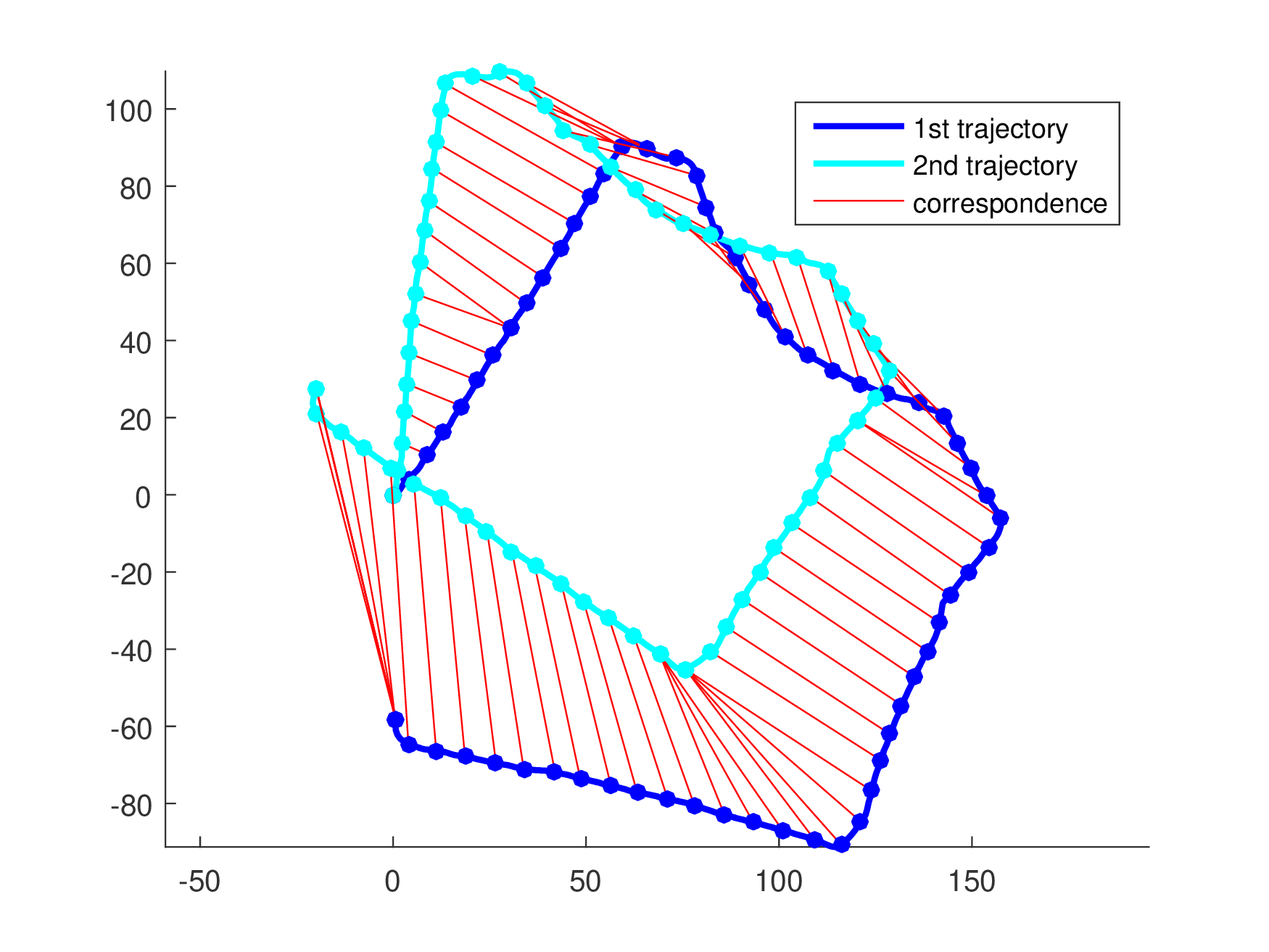}}
\caption{DTW-metric}
\end{subfigure}\\
\begin{subfigure}[b]{\linewidth}
\centering%
\centerline{\includegraphics[width=9cm]{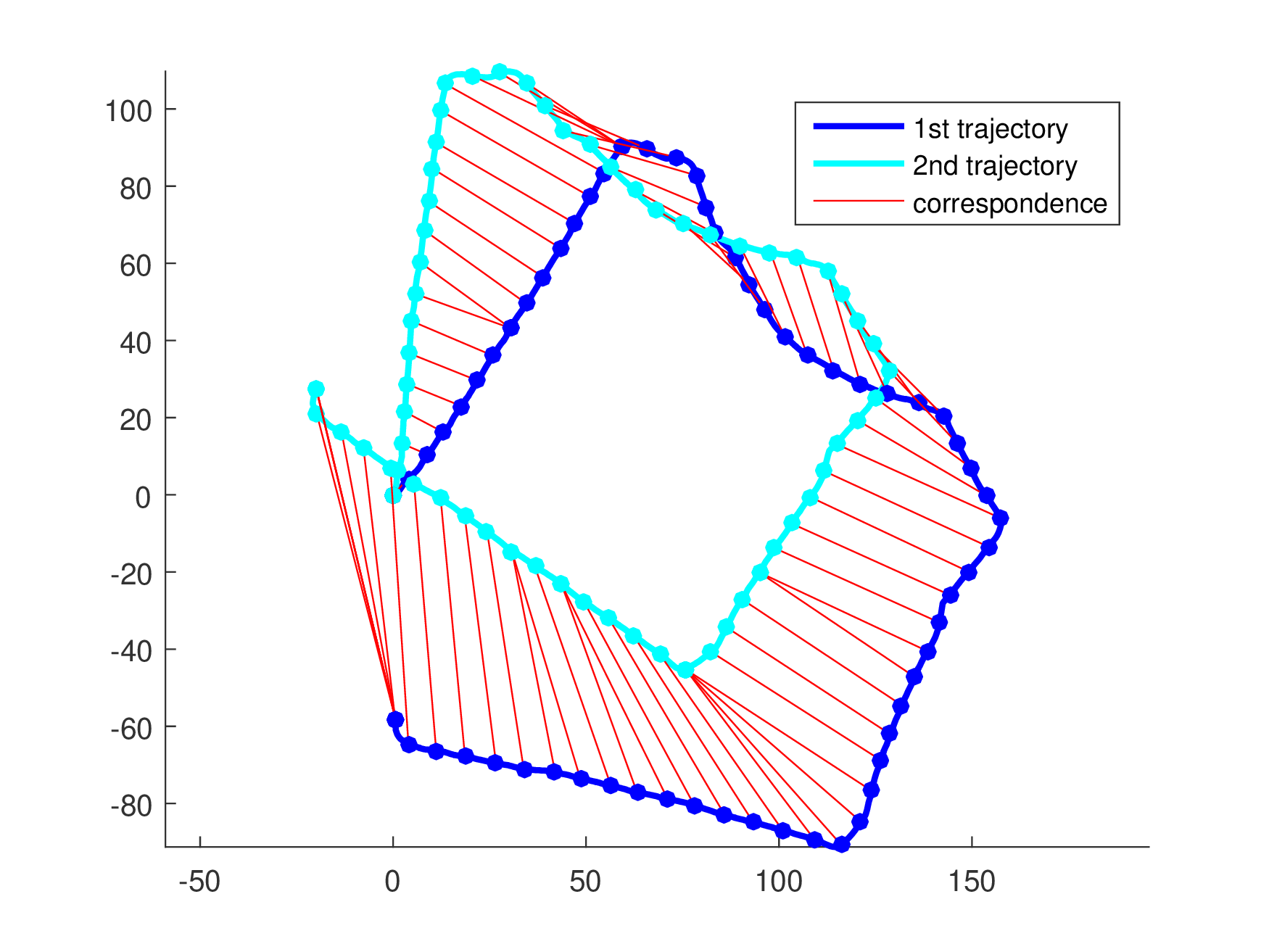}}
\caption{Fr\'echet metric}
\end{subfigure}
\captionsetup{justification=centering}
\caption{Comparison of metrics with the use of additional constraint.}\label{8-158-3}
\end{figure}

It should be noted that the most simple algorithms for calculation of the mentioned metrics use a method of dynamic programming, in which distances in all pairs of points are calculated.
However, the approach with an additional constraint reduces the algorithm time complexity as well as the space complexity from $O(nm)$ to $O(w \cdot \max(n,m))$.

\subsection{Path beam averaging}

The results of each introduced algorithm also include a sequence of index pairs that determines the mapping of one curve to another. If we take the middle point of every segment joining a pair of ``similar'' points, we will obtain a certain new curve.
An analysis of such a method of creating the ``average'' trajectory using the DTW algorithm is presented in~[10].
However, that method of averaging loses information about time of measurements since the points mapped to each other can be obtained at different moments. If that information is important, points with equal sequence numbers should be matched.

\section{Algorithm for two IMUs}

In the middle of a step the position of a moving foot is supposed to be approximately equal to a position of the otherot foot. According to that idea, the additional observations may be passed to the Kalman filter. Thus, it is suggested to calculate trajectories of different legs side by side.

\begin{algorithm}\label{both_filter}
\SetAlgoLined
\DontPrintSemicolon
\SetNoFillComment
$cur \gets 2$ \tcp*{begin computation with the second leg}
$\vect{pos} \gets \tilde{\vect{x}}_{k_1}^1$ \tcp*{current position of the first leg}
\While{$k_1 < N_1\ || \ k_2 < N_2$}{
    \tcc*[l]{select data for the current leg}
    \lIf{cur = 1}{
        $\tilde{\vect{x}}_{n-1} \gets \tilde{\vect{x}}_{n-1}^1$, $k \gets k^1$, $\tilde{\vect{f}}_n \gets \tilde{\vect{f}}_n^1$, $\tilde{\vect{w}}_n \gets \tilde{\vect{w}}_n^1$
    }
    \lElse {
        $\tilde{\vect{x}}_{n-1} \gets \tilde{\vect{x}}_{n-1}^2$, $k \gets k^2$, $\tilde{\vect{f}}_n \gets \tilde{\vect{f}}_n^2$, $\tilde{\vect{w}}_n \gets \tilde{\vect{w}}_n^2$
    }
    \ \\
    \tcc*[l]{process movement of the current leg}
    $step\_len$ = count\_step\_length($k$, $\{\tilde{\vect{f}}_i, \tilde{\vect{w}}_i\}$) \\
    \For{$n = k+1$ \KwTo $k+step\_len$}{
        $[\tilde{\vect{x}}_n, \delta \tilde{\vect{x}}_{n|n-1}, \vect{P}_{n|n-1}] \gets$
        predict($\tilde{\vect{x}}_{n-1}, \tilde{\vect{f}}_n, \tilde{\vect{w}}_n$)
        \If{$n = k + step\_len / 2$}{
            \tcc*[l]{correction step of Kalman filter using current position of another leg}
            \begin{math}
            \vect{H} \gets
            \begin{bmatrix}
            \vect{I}_{2 \times 2} & \vect{O}_{2 \times 7} \\
            \end{bmatrix}
            \end{math} \\
            $\vect{K}_n \gets \vect{P}_{n|n-1} \vect{H}^T \left( \vect{H}\vect{P}_{n|n-1} \vect{H}^T + \vect{R}''' \right)^{-1}$ \\
            \ \\
            \tcc*[l]{select 2 coordinates}
            $\tilde{\vect{x}}_n^{1,2} \gets \vect{H} \tilde{\vect{x}}_n,\ \delta \tilde{\vect{x}}_{n|n-1}^{1,2} \gets \vect{H} \delta \tilde{\vect{x}}_{n|n-1}$\\
            \ \\
            $\delta \tilde{\vect{x}}_{n|n} \gets \delta \tilde{\vect{x}}_{n|n-1} - \vect{K}_n (\vect{pos} + \delta \tilde{\vect{x}}_{n|n-1}^{1,2} - \tilde{\vect{x}}_n^{1,2})$ \\
            $\vect{P}_{n|n} \gets \left( \vect{I}_{9 \times 9} - \vect{K}_n \vect{H} \right) \vect{P}_{n|n-1}$ \\
        }
    }
    $n \gets k + step\_len + 1$ \\
    \ \\
    \tcc*[l]{stationary position of the current leg}
    \While{$T\left(\{\tilde{\vect{f}}_i, \tilde{\vect{w}}_i\}_{W_n}\right) < \gamma$}{
        $[\tilde{\vect{x}}_n, \delta \tilde{\vect{x}}_{n|n-1}, \vect{P}_{n|n-1}] \gets$ predict($\tilde{\vect{x}}_{n-1}, \tilde{\vect{f}}_n, \tilde{\vect{w}}_n$) \\
        \uIf{\normalfont{standstill($n$)} = true}{
            \begin{math}
            \vect{H} \gets
            \begin{bmatrix}
            \vect{I}_{3 \times 3} & \vect{O}_{3 \times 3} & \vect{O}_{3 \times 3} \\
            \vect{O}_{3 \times 3} & \vect{I}_{3 \times 3} & \vect{O}_{3 \times 3}
            \end{bmatrix},
            \end{math}
            $\vect{R} \gets \vect{R}'$
        }
        \Else{
            \begin{math}
            \vect{H} \gets
            \begin{bmatrix}
            \vect{O}_{3 \times 3} & \vect{I}_{3 \times 3} & \vect{O}_{3 \times 3}
            \end{bmatrix},
            \end{math}
            $\vect{R} \gets \vect{R}''$
        }
        $[\delta \tilde{\vect{x}}_{n|n}, \vect{P}_{n|n}] \gets$ correct($\vect{H}, \vect{R}, \vect{P}_{n|n-1}, \tilde{\vect{x}}_n, \delta \tilde{\vect{x}}_{n|n-1}$) \\
        $n \gets n + 1$
    }
    \ \\
    \tcc*[l]{save current position for the following observation}
    $\vect{pos} \gets \tilde{\vect{x}}_n^{1,2}$ \\
    \ \\
    \tcc*[l]{save updated data}
    \uIf{cur = 1}{
        $\tilde{\vect{x}}_n^1 \gets \tilde{\vect{x}}_n$, $\delta \tilde{\vect{x}}_{n|n}^1 \gets \delta \tilde{\vect{x}}_{n|n}$, $\vect{P}_{n|n}^1 \gets \vect{P}_{n|n}$, $k^1 \gets n$ \\
        $cur \gets 2$
    }
    \Else {
        $\tilde{\vect{x}}_n^2 \gets \tilde{\vect{x}}_n$, $\delta \tilde{\vect{x}}_{n|n}^2 \gets \delta \tilde{\vect{x}}_{n|n}$, $\vect{P}_{n|n}^2 \gets \vect{P}_{n|n}$, $k^2 \gets n$ \\
        $cur \gets 1$
    }
}
\caption{Algorithm that uses data from both IMUs.}
\end{algorithm}

The algorithm consists of the following steps:
\begin{enumerate}
    \item Both trajectories are calculated separately using the filter described in Algorithm~3. Then trajectories are overlapped to determine the initial difference between the yaw angles. For that purpose, the DTW distance is minimized by selecting the initial angles.
    \item A leg that makes the first step is determined. It is supposed that the first step is made from the initial standstill position and is shorter than the first step of another leg.
    \item
    Two ``fused'' trajectories are сomputed using data from both sensors.
    \item RTS smoothing is applied to both calculated trajectories (see Algorithm~3).
    \item The obtained trajectories are optimally overlapped once again using brute-force angle search. Then the average trajectory is calculated with use of the DTW-algorithm.
\end{enumerate}

Step~3 is described in detail in Algorithm~4 (variables with upper index 1 correspond to a leg (IMU) that makes the first step; upper index 2 stands for the other IMU).

Functions \textbf{predict} and \textbf{correct} stand for prediction and correction steps of the Kalman filter (see Algorithm~3).

\section{Results}

Fig.~6-7 contain trajectories obtained with an algorithm which does not take the final observation into account (nevertheless, the smoothing is applied). Also, trajectories reconstructed with Algorithms~3~and~4 are presented there (trajectories of left and right legs are marked in yellow and cyan colors respectively; red color stands for the final generalized trajectories). The DTW-distances between corresponding trajectories of left and right legs are indicated as well. Trajectories in Fig.~7 correspond to a double pass back and forth along a curved corridor.

\begin{table}[H]
\begin{center}
\begin{tabular}{|c|c|c|c|c|}
\hline
N & Alg.~[1] & Alg.~3 & Alg.~4 & Duration, s  \\
\hline
1 & 4.460 & 0.292 & 0.252 & 93,5 \\
\hline
2 & 20.430 & 1.317 & 0.723 & 216.7 \\
\hline
3 & 14.981 & 3.206 & 0.939 & 245.5 \\
\hline
4 & 16.565 & 1.437 & 1.042 & 250,9 \\
\hline
5 & 6.690 & 1.951 & 1.491 & 332.8 \\
\hline
6 & 45.630 & 17.917 & 1.546 & 630.1 \\
\hline
\end{tabular}
\caption{Results of the presented algorithms.}
\label{table_res}
\end{center}
\end{table}

The figures illustrate the proximity of resconstructed trajectories of the INS attached to different legs of a person. 
Moreover, the presented trajectories reproduce the metric characteristics of rooms in which the experiments were conducted and it further confirms the correctness of the restoration, allowing the use of the proposed algorithms when more reliable reference information is not available.

Table~I shows the duration times of several experiments and values of DTW metric, obtained as results of the presented algorithms. Trajectories~1~and~2 correspond to a single pass back and forth along a straight line, trajectory 3 was obtained as a result of a double pass around the perimeter of the room. The trajectories~4~and~5 correspond to a single pass back and forth along a L-shaped corridor, trajectory 6 corresponds to a double pass along it.

The values of DTW metric do not give a direct answer to the question about the metric difference of the reconstructed and real trajectories, but they allow us to state that the proposed algorithms reconstruct real trajectories with significantly reduced error compared to trajectories reconstructed according to a single INS.

\begin{figure}[H]
\begin{subfigure}[b]{\linewidth}
\centering%
\centerline{\includegraphics[width=9cm]{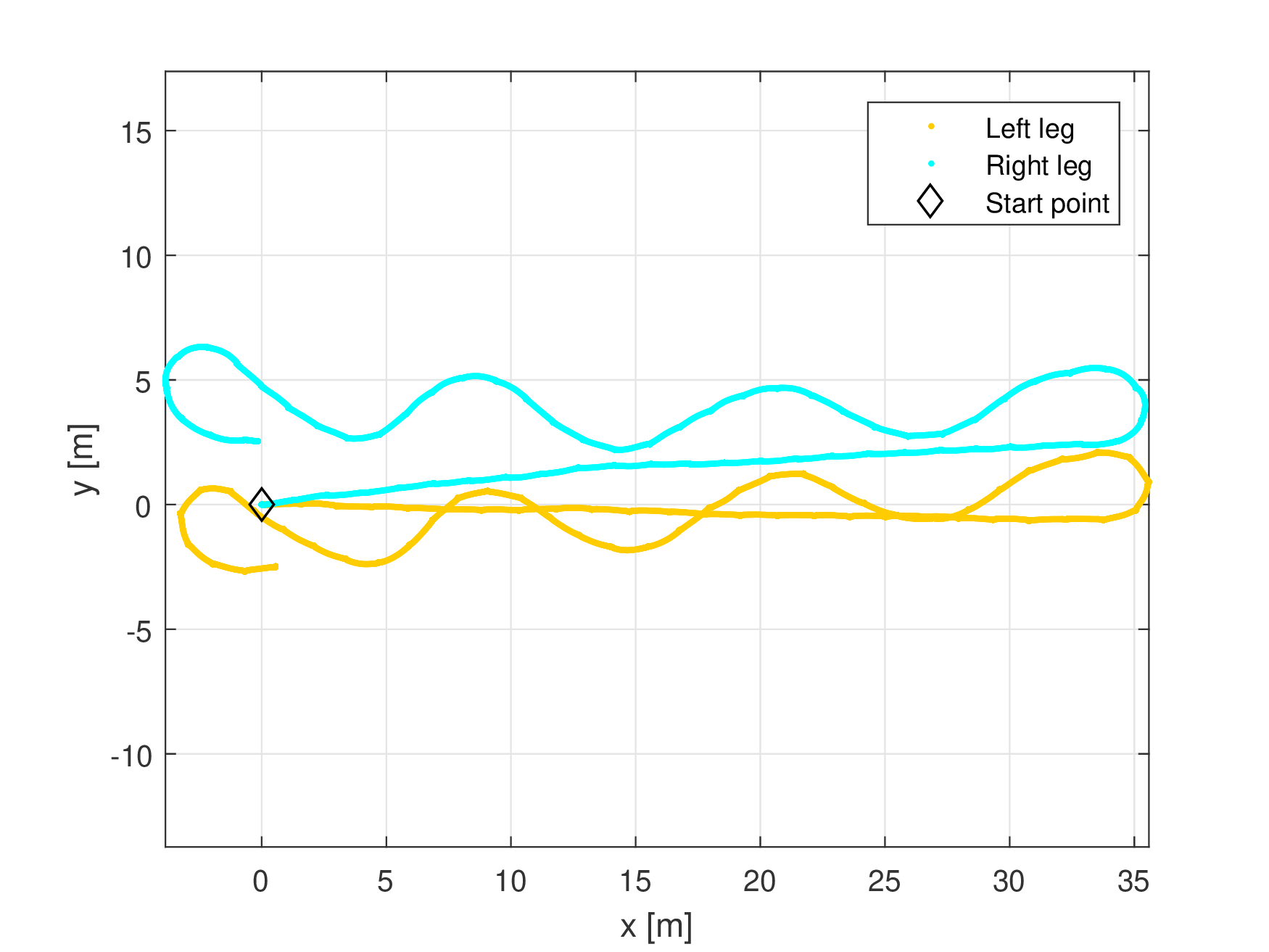}}
\caption{Alg.~[1], \hbox{$d = 2.949$}.}
\end{subfigure}\\
\begin{subfigure}[b]{\linewidth}
\centering%
\centerline{\includegraphics[width=9cm]{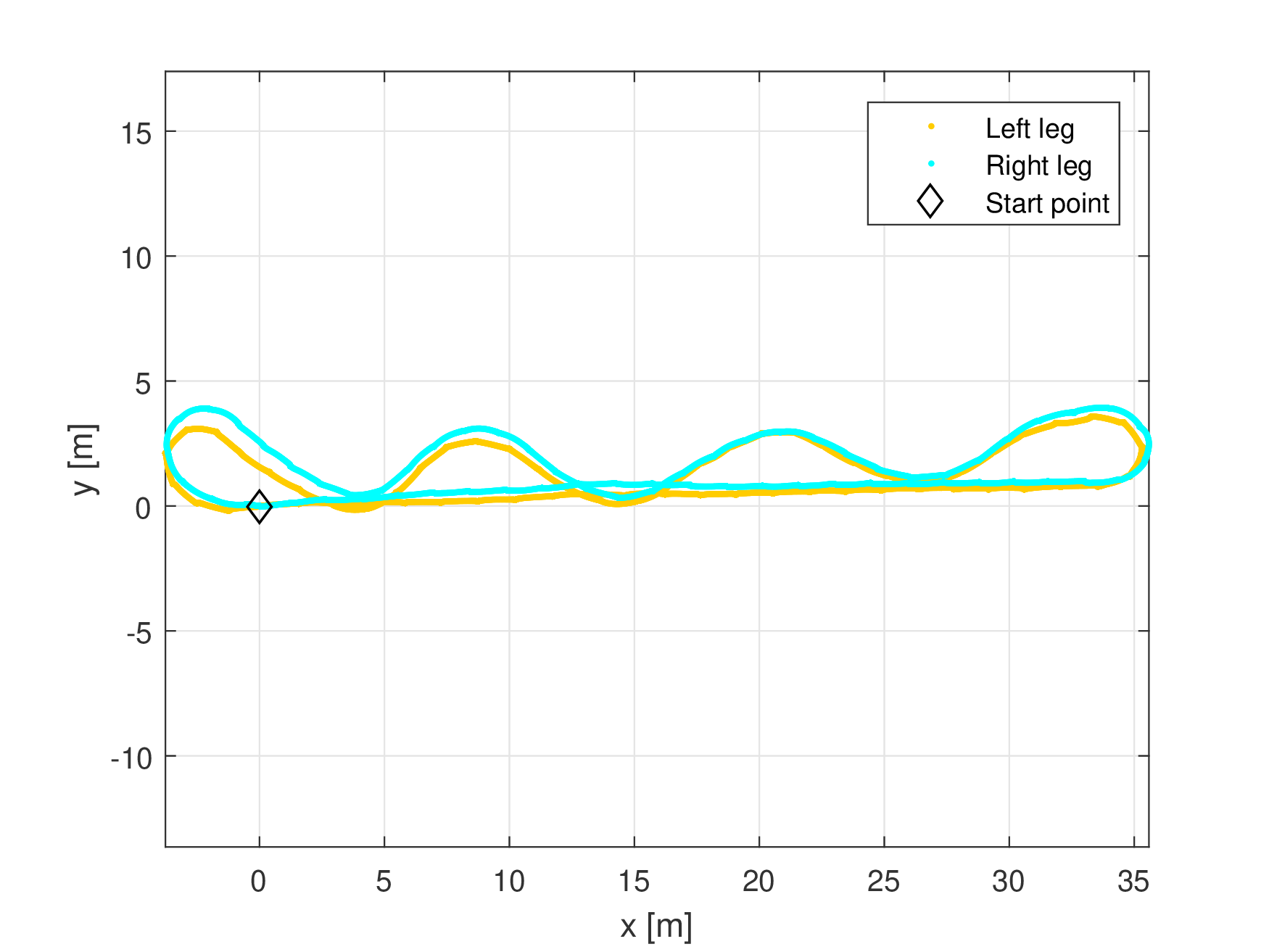}}
\caption{Alg.~3, \hbox{$d = 0.241$}.}
\end{subfigure}
\begin{subfigure}[b]{\linewidth}
\centering%
\centerline{\includegraphics[width=9cm]{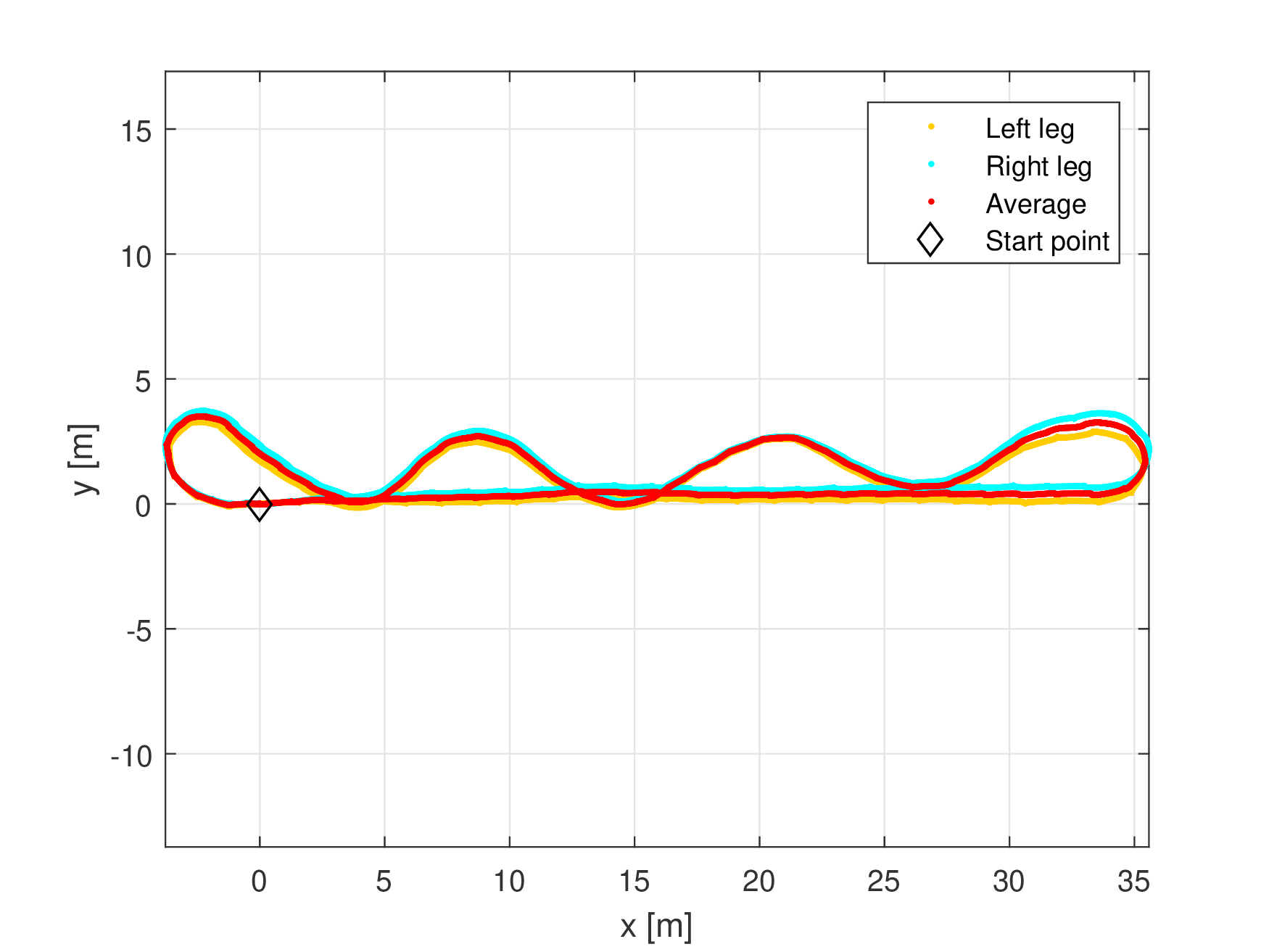}}
\caption{Alg.~4, \hbox{$d = 0.283$}.}
\end{subfigure}
\captionsetup{justification=centering}
\caption{Comparison of algorithms.}\label{traj_1-12}
\end{figure}

\begin{figure}
\begin{subfigure}[b]{\linewidth}
\centering%
\centerline{\includegraphics[width=9cm]{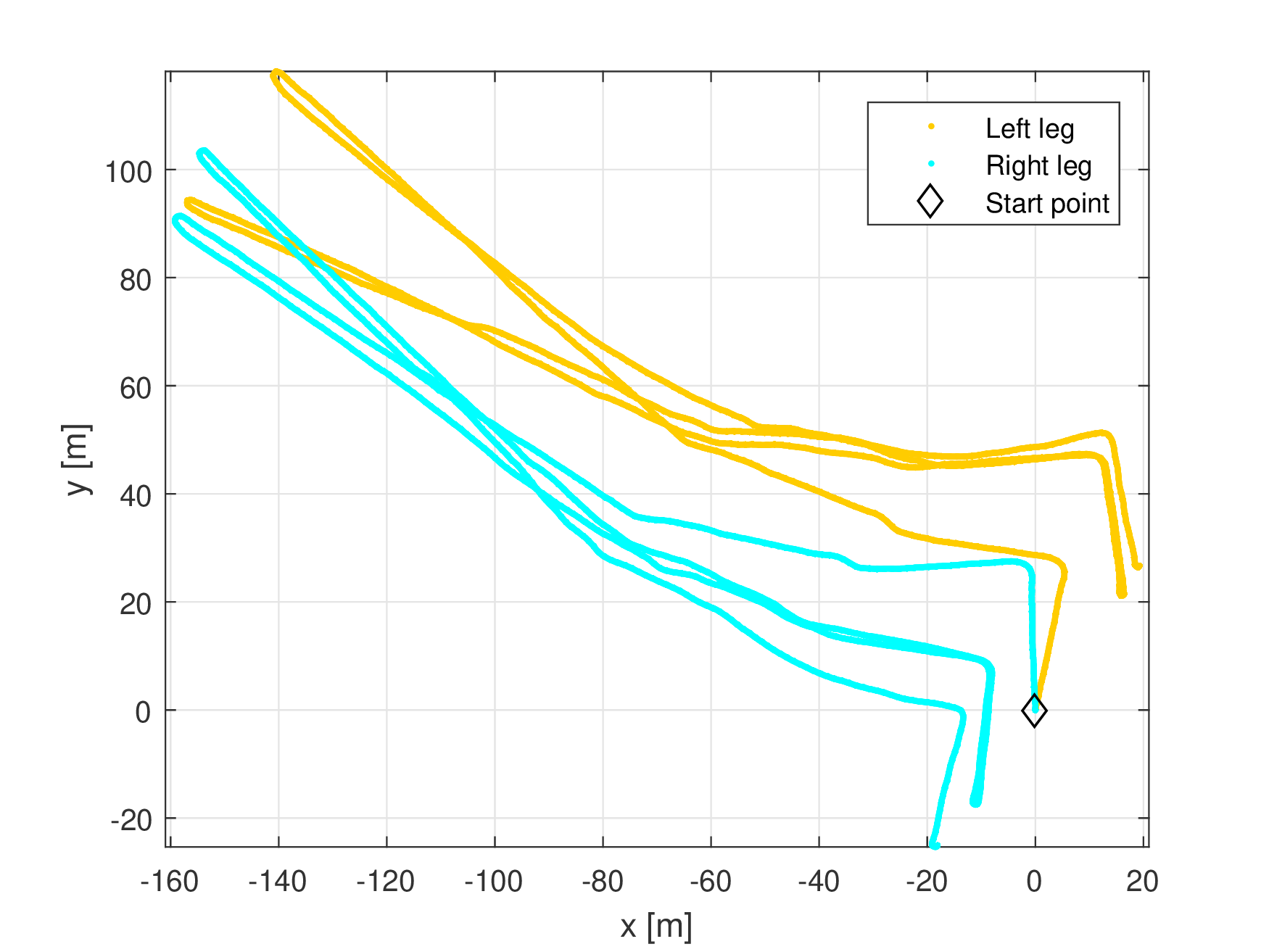}}
\caption{Alg.~[1], \hbox{$d = 25.862$}.}
\end{subfigure}\\
\begin{subfigure}[b]{\linewidth}
\centering%
\centerline{\includegraphics[width=9cm]{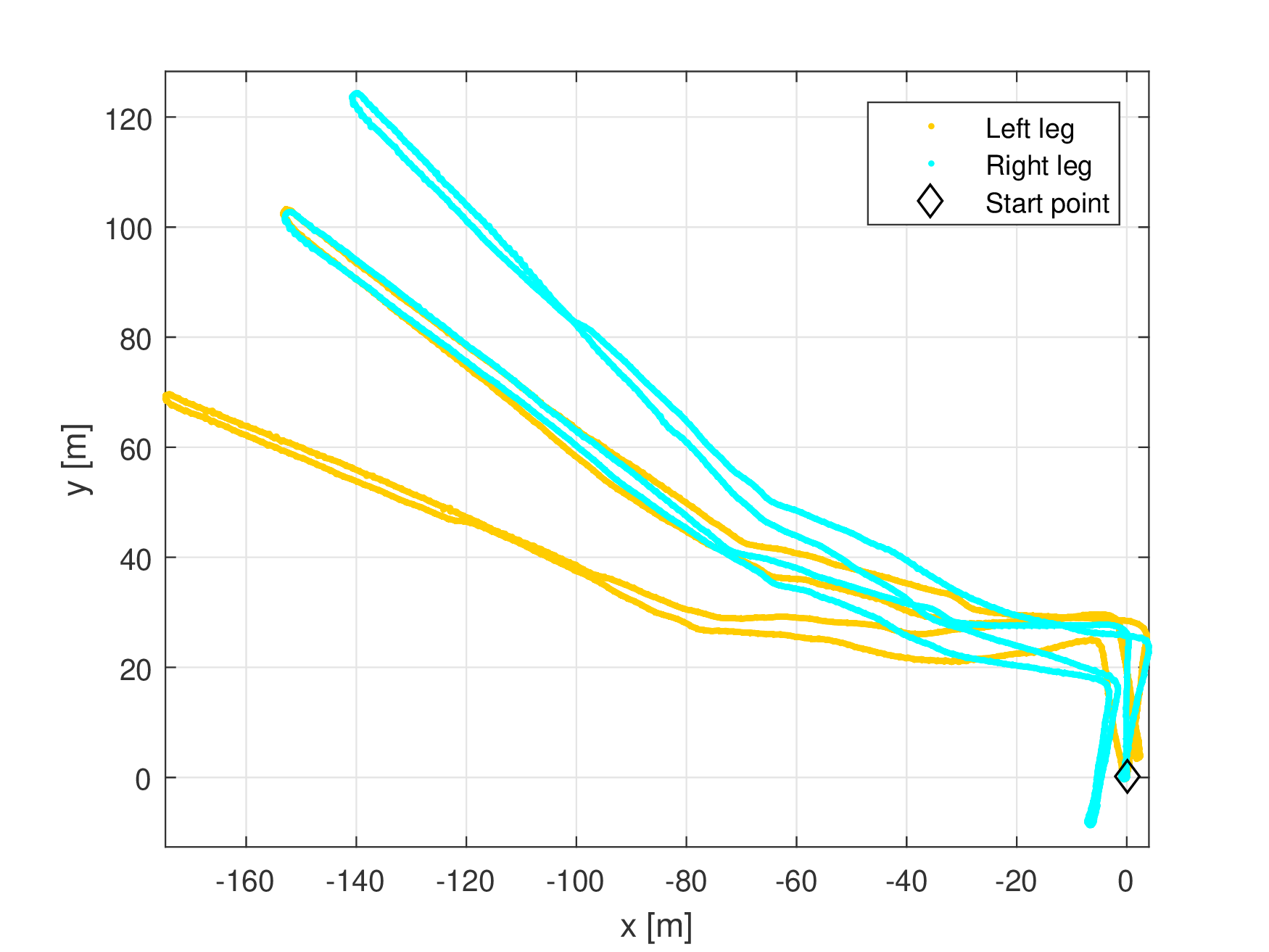}}
\caption{Alg.~3, \hbox{$d = 13.883$}.}
\end{subfigure}
\begin{subfigure}[b]{\linewidth}
\centering%
\centerline{\includegraphics[width=9cm]{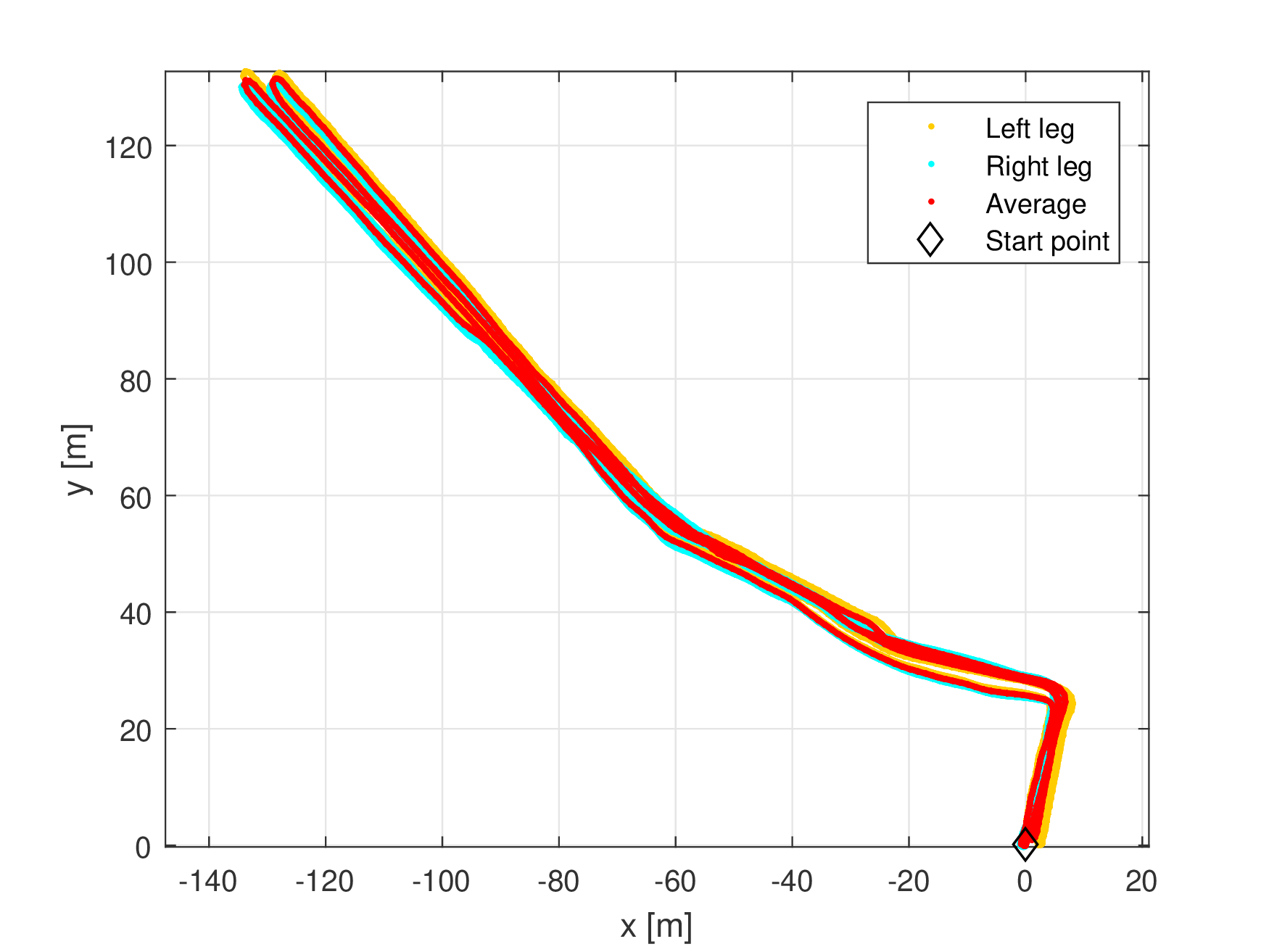}}
\caption{Alg.~4, \hbox{$d = 1.354$}.}
\end{subfigure}
\captionsetup{justification=centering}
\caption{Comparison of algorithms.}\label{traj_2-12}
\end{figure}

\section{Conclusion}

The paper contains description of the algorithms for reconstruction of close-loop trajectories based on information about acceleration and angular velocity (for only one IMU and for IMUs on both legs). The corresponding pseudocode is presented. Two ways of comparing the obtained trajectories are proposed, their advantages and disadvantages are considered, and a method of optimizing the computation time is specified. In addition, a method for constructing a single combined trajectory based on measurements from the IMUs installed on each of two feet is proposed. Both proposed algorithms were tested on real data and thus demonstrated their efficiency as a tool for obtaining reference paths when more accurate reference information is not available.


\end{document}